\documentclass[aps,prxquantum,reprint,superscriptaddress,nofootinbib,floatfix]{revtex4-2}
\usepackage{graphicx}
\usepackage{amsmath,amssymb,bm}
\usepackage{xcolor}
\usepackage{tikz}
\usetikzlibrary{decorations.markings,shapes.geometric,arrows.meta}
\usepackage[colorlinks=true,citecolor=blue,linkcolor=red,urlcolor=blue]{hyperref}
\usepackage{placeins}
\renewcommand{\topfraction}{0.9}
\renewcommand{\bottomfraction}{0.7}
\renewcommand{\textfraction}{0.08}
\renewcommand{\floatpagefraction}{0.75}

\newcommand{\up}{\uparrow}
\newcommand{\down}{\downarrow}
\definecolor{okvermillion}{HTML}{D55E00}
\definecolor{okblue}{HTML}{0072B2}
\definecolor{okgrey}{HTML}{7F7F7F}
\tikzset{
  op/.style={draw, fill=okblue!20, minimum size=7pt, inner sep=0pt},
  gate/.style={draw, fill=okgrey!25, minimum size=7pt, inner sep=0pt},
  rho/.style={draw, fill=okvermillion!25, minimum size=7pt, inner sep=0pt, circle},
  proj/.style={draw, fill=okgrey!40, isosceles triangle, inner sep=1pt,
               minimum size=5pt},
  leg/.style={thick},
  msg/.style={draw, fill=black, circle, minimum size=5pt, inner sep=0pt},
  fan/.style={draw, fill=okgrey!40, thick},
}

\makeatletter
\def\@hangfrom@section#1#2#3{\@hangfrom{#1#2}#3}
\def\@hangfroms@section#1#2{#1#2}
\makeatother

\begin{document}

\title{Fast classical simulation of `Fast, accurate, high-resolution simulation of large-scale Fermi-Hubbard models on a digital quantum processor'}
\author{Xiao-Yu Ouyang}
\affiliation{Division of Chemistry and Chemical Engineering,
  California Institute of Technology, Pasadena, California 91125, USA}
\affiliation{Marcus Center for Theoretical Chemistry,
  California Institute of Technology, Pasadena, California 91125, USA}
\author{Runze Chi}\affiliation{Division of Chemistry and Chemical Engineering,
  California Institute of Technology, Pasadena, California 91125, USA}
\affiliation{Marcus Center for Theoretical Chemistry,
  California Institute of Technology, Pasadena, California 91125, USA}
\author{Garnet Kin-Lic Chan}
\affiliation{Division of Chemistry and Chemical Engineering,
  California Institute of Technology, Pasadena, California 91125, USA}
\affiliation{Marcus Center for Theoretical Chemistry,
  California Institute of Technology, Pasadena, California 91125, USA}

\affiliation{Institute for Quantum Information and Matter, California Institute of Technology, Pasadena, California 91125, USA}
\date{\today}

\begin{abstract}
We study the N\'{e}el quench dynamics of a 1D Fermi-Hubbard model which has recently been simulated on quantum hardware~\cite{hartnettFastAccurateHighresolution2026a}. We demonstrate that the set of 7260 observable trajectories measured in the quantum experiment can be obtained more quickly and accurately through classical tensor network simulation using modest computation. Our result relies on transverse tensor network contraction, where a bond dimension of 32 is already sufficient to reproduce the quantum experiment. We further extend the converged observable trajectories to longer times than in the hardware simulation and in other recent classical simulations.  
\end{abstract}

\maketitle
\section{Introduction}

Quantum computers are theoretically efficient simulators of quantum dynamics~\cite{feynmanSimulatingPhysicsComputers1982,lloydUniversalQuantumSimulators1996}, thus quantum simulation problems of a dynamical nature have been anticipated to be ones where quantum advantage can be readily observed and claimed~\cite{kim2023evidence,king2025beyond,haghshenas2026digital,google2025observation}. Assessing the results of these quantum experiments has involved a fruitful dialog with classical simulations, with a number of  claims challenged by improved classical techniques~\cite{beguvsic2024fast,tindall2024efficient,tindall2026dynamics}.

Recently, the short-time quench dynamics of the 1D Fermi-Hubbard model~\cite{hartnettFastAccurateHighresolution2026a} has been proposed as a problem where a time-to-solution `practical quantum advantage' has been obtained. 
The setting for this statement is a Trotterized circuit dynamics of a 60-site (120 qubit) Hubbard problem, with the readout being 7260 observables (one- and two-point density correlators), with an approximate accuracy of 1\% up to $t=5.2$ (time units of inverse hopping $1/t_h$), quenching from an initial N\'{e}el state. This task took 166s (265s with decay recovery and measurement mitigation) on an IBM quantum processor, with other classical processing timecosts neglected.
The initial proposed advantage was $\sim 3000\times$ speedup~\cite{hartnettFastAccurateHighresolution2026av1}, subsequently revised to a $\sim 500\times$ speedup~\cite{hartnettFastAccurateHighresolution2026a}, over available classical methods to compute all observables, based on 
a time-dependent variational principle (TDVP) matrix product state simulation~\cite{dorando2009analytic,haegeman2011time,nakatani2014linear}
 with bond dimension $\chi=4096$, executed on 32 virtual CPUs. The same work also benchmarked classical methods that could compute individual observables more quickly, such as matrix product operator evolution~\cite{zwolakMixedStateDynamics2004,prosenOperatorSpaceEntanglement2007,hartmannDensityMatrixRenormalization2009}, Pauli propagation~\cite{rudolph2023classical,beguvsic2024fast,beguvsic2025simulating,shao2024simulating}, and Majorana propagation~\cite{miller2025simulation}, with the best per-observable result at the same accuracy as the quantum simulation in Ref.~\cite{hartnettFastAccurateHighresolution2026a} coming from MPO simulation at $\sim 1000$s. While such calculations could be parallelized to obtain all observables in the same time,  Ref.~\cite{hartnettFastAccurateHighresolution2026a} suggested that this would be impractical, given there were 7260 of them. Finally, a recent work using a GPU based TDVP implementation with symmetry~\cite{rauschPushingClassicalFrontier2026} further reduced the classical cost, leaving only a $36\times$ walltime quantum advantage.

Here we show that the same quantum experiment can actually be simulated classically \emph{faster than the quantum processing time}, using algorithms implemented within the \textsf{Quimb} package~\cite{grayQuimbPythonPackage2018}.
Using transverse tensor contraction approximation whose accuracy depends on temporal entanglement entropy, we simulate a single observable in 51s on a single GPU, and, through tensor network reuse, all 7260 observables can be simulated on at most 120~GPUs in less than 166s.
The latter computational resource is readily available through cloud computing and corresponds to a total of $\sim 30$ USD of cloud time with standard compute instances. 
We also find that using the improved MPO simulation algorithm in \textsf{Quimb}, each observable can be computed in only 12s on a single GPU.
The improvements over prior results arise from a mixture of technical contributions in \textsf{Quimb} (GPU implementation, stochastic randomized compression and environment-dependent compression, and tensor network reuse) and fundamental reasons (temporal and operator entanglement versus spatial wavefunction entanglement). 
We further demonstrate that using these methods, with modest bond dimensions ($\chi=32$ for transverse contraction) we can extend the simulation time beyond the quantum experiment and the TDVP simulation of Ref. \cite{rauschPushingClassicalFrontier2026} with similar or better accuracy.
\FloatBarrier
\section{Setup and Method}
\label{sec:method}
The Trotterized quench dynamics of the 1D Fermi Hubbard model starts with the Hamiltonian
\begin{equation}
    H=-t_h \sum_{i=0}^{L-2}\sum_{\sigma}\left(c_{i\sigma}^{\dagger}c_{i+1,\sigma}+\text{h.c.}\right)+U\sum_{i=0}^{L-1}n_{i\uparrow}n_{i\downarrow}.
\end{equation}
$c_{i \sigma}^{\dagger}$ is the creation operator of $\operatorname{spin} \sigma \in\{\uparrow, \downarrow\}$ on site $i$, and $n_{i \sigma}=c_{i \sigma}^{\dagger} c_{i \sigma}$ is the number operator on site $i$ with spin $\sigma$. The physical sites are indexed by $i=0,\ldots,L-1$. $t_h$ is the hopping and $U$ is the interaction term. To simulate the quantum experiment, we set the 1D chain length to be $L=60$ (120 qubits), $t_h=1$ and $U=-2$.

Ref.~\cite{hartnettFastAccurateHighresolution2026a} used a Trotterized approximation of $e^{-iHt}$, decomposing the Hamiltonian as $H=H_A+H_B+h$
\begin{equation}
    \begin{aligned}
H_A
&=
-t_h\left(
\sum_{i\ \mathrm{even}}
c_{i\uparrow}^{\dagger}c_{i+1,\uparrow}
+
\sum_{i\ \mathrm{odd}}
c_{i\downarrow}^{\dagger}c_{i+1,\downarrow}
+\mathrm{h.c.}
\right),\\
H_B
&=
-t_h\left(
\sum_{i\ \mathrm{even}}
c_{i\downarrow}^{\dagger}c_{i+1,\downarrow}
+
\sum_{i\ \mathrm{odd}}
c_{i\uparrow}^{\dagger}c_{i+1,\uparrow}
+\mathrm{h.c.}
\right)
,\\
h
&=U\sum_{i=0}^{L-1} n_{i\uparrow}n_{i\downarrow}.
\end{aligned}
\end{equation}
The time evolution unitary operator was approximated using alternating Trotter gates
\begin{equation}
\label{eq:trotter-gates}
\begin{aligned}
 U_A(dt)
&=
e^{-i\,dt H_B}
e^{-i\,dt h}
e^{-i\,dt H_A},\\
 U_B(dt)
&=
e^{-i\,dt H_A}
e^{-i\,dt h}
e^{-i\,dt H_B}.
\end{aligned}
\end{equation}
The quantum experiment used the sequence $U_B(dt)U_A(dt)$ to implement a second-order Trotter approximation with $2dt$ as the Trotter step~\cite{hartnettFastAccurateHighresolution2026a}, choosing $dt=0.2 t_h^{-1}$. 

The initial state in \cite{hartnettFastAccurateHighresolution2026a} was prepared as the half-filled N\'{e}el product state $|\psi_0\rangle=|\downarrow\uparrow\downarrow\uparrow\cdots\rangle$. 
Using 20000 measurement shots at each time point, the experiment obtained the occupation number bitstrings, from which the local density and two-site correlations on all the sites were computed. 

In a classical simulation, the task of carrying out the time evolution and computing the observables can be formulated as a tensor network (TN) contraction. Representing each layer of Trotterized gates as a matrix product operator, and grouping $\up, \down$ spins onto a single physical site, we obtain a tensor network as shown in Fig.~\ref{fig:tn-contraction}, where both the matrix product operator bond dimension $D$ and the physical bond dimension $d$ is 4. We used the same mirrored circuit construction as in the quantum experiment.

The classical simulation task is then to contract this tensor network for the observables. For a TN of this size, exact contraction (which depends exponentially on the size of the tensor network) is not possible, thus the two choices in designing the tensor network simulation are (i) the order in which the tensors are contracted and (ii) how the resulting intermediate tensor networks are compressed. We consider two types of approximate tensor network simulation:

\begin{figure}
    \centering
    \includegraphics[width=1.05\linewidth]{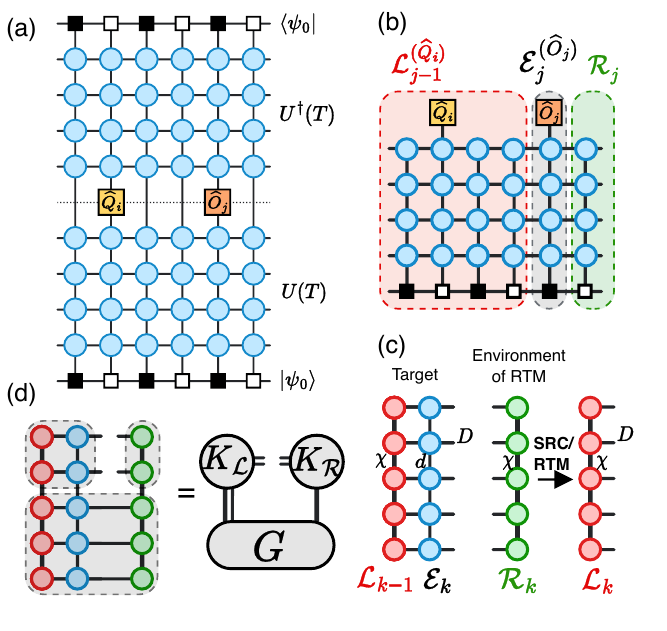}
    \caption{Transverse contraction of the spacetime tensor network (TN) for a 1D fermionic chain.
    (a) TN diagram of a nonlocal two-point observable expecation value \(\langle \hat{Q}_i(t) \hat{O}_j(t)\rangle = \langle{\psi_0}| U^\dagger(T) \hat{Q}_i \hat{O}_j U(T)|{\psi_0}\rangle \). 
    (b) The spacetime TN after folding along the dashed line in (a), here the grey shaded part is the temporal MPO (tMPO) \(\mathcal{E}_j^{(\hat{O}_j)} \) on site $j$ that contains operator $\hat{O}_j$, the red shaded part is the left environment $\mathcal{L}_{j-1}^{(\hat{Q}_i)}$ of site $j$ with operator $\hat{Q}_i$ included, and the green shaded part is the right environment $\mathcal{R}_{j}$.
    (c) Compression of the temporal MPS (tMPS) +  tMPO target $\mathcal{L}_{k-1} \cdot \mathcal{E}_k$ at site $k$ using its environment tMPS $\mathcal{R}_k$ (we either use successive randomized compression (SRC) without environment information, or reduced transition matrix (RTM) compression that uses the environment information). $\chi$ is the virtual bond dimension of the tMPS $\mathcal{L}_{k-1}$, $d$ the virtual bond dimension of the tMPO $\mathcal{E}_k$, $D$ is the physical bond dimension of each tMPS and tMPO. The outcome of compression is the tMPS $\mathcal{L}_k$ with bond dimension $\chi$.
    (d) Diagram of a reduced transition matrix to be truncated, which can be represented as ${K}_\mathcal{L} \cdot G \cdot {K}_\mathcal{R}$ by contracting each grey shaded part.
    }.

    \label{fig:tn-contraction}
\end{figure}

\begin{figure}
    \centering
    \includegraphics[width=1\linewidth]{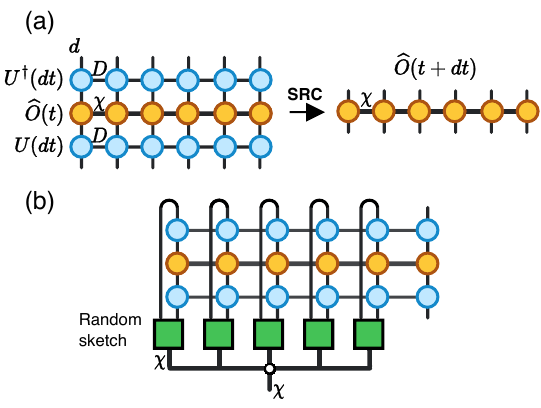}
    \caption{MPO Heisenberg evolution of an observable in a 1D chain.
    (a) TN diagram to obtain MPO $\hat{O}(t+dt)$ from $U^\dagger(dt)\hat{O}(t) U(dt)$. $\chi$ is the virtual bond dimension of the operator MPO, $D$ is the virtual bond dimension of the unitary MPO, and $d$ is the physical bond dimension of each MPO.
    (b) Successive randomized compression for the MPO-MPOs product on the most right site, starting from applying random sketched matrices (green) on the open indices.
    }
    \label{fig:mpo-heisenberg}
\end{figure}

\noindent \emph{Transverse contraction with SRC and RTM}. 
The main strategy considered in this work is transverse contraction, i.e. the network in Fig.~\ref{fig:tn-contraction}(a) is contracted along the spatial rather than temporal direction~\cite{banulsMatrixProductStates2009,mullerHermesTensorNetwork2012,hastingsConnectingEntanglementTime2015,friasPerezLightConeTensor2022}.
As is common in temporal contraction, we first fold the network in Fig.~\ref{fig:tn-contraction}(b) along the dashed line in Fig.~\ref{fig:tn-contraction}(a), then contract from the left boundary through site $j-1$ (creating the left environment $\mathcal{L}_{j-1}$ in Fig.~\ref{fig:tn-contraction}(b)) and from the right boundary through site $j+1$, creating $\mathcal{R}_{j}$, propagating left and right \emph{temporal} MPSs (tMPSs) of bond dimension $\chi$. The first left tMPS $\mathcal{L}_0$ is the first temporal column.
This temporal boundary-MPS construction is closely related to tensor network representations of many-body influence matrices and influence functionals~\cite{leroseInfluenceMatrixApproach2021,sonnerInfluenceFunctionalManybody2021,yeConstructingTensorNetwork2021,parkSimulatingQuantumDynamics2025}.
As each column is contracted into $\mathcal{L}$ or $\mathcal{R}$, it is compressed. 
Here we use the the stochastic randomized compression algorithm~\cite{camanoSuccessiveRandomizedCompression2026a}, as well as reduced transition matrix (RTM) compression~\cite{carignanoTemporalEntropyComplexity2024a}, the latter of which uses the  temporal environment of uncontracted columns 
to improve the accuracy of the temporal MPS contraction.
For example, in Fig.~\ref{fig:tn-contraction}(c), when contracting $\mathcal{E}_k$ into $\mathcal{L}_{k-1}$, the RTM compression is informed by the `environment' of columns on the right $\mathcal{R}_k$~\cite{hastingsConnectingEntanglementTime2015,cerezoSpatiotemporalTensorNetwork2025,carignanoITransverseLibrary2026}. 

We now briefly describe an improvement we use over the method originally proposed in Ref.~\cite{carignanoITransverseLibrary2026}.
As illustrated in Fig.~\ref{fig:tn-contraction}(c,d), given the tMPS-tMPO product and the environment tMPS, to compress the selected temporal cut, we first obtain the reduced transition matrix $\rho$ by creating the partial trace of the transition matrix $\mathcal{L} \otimes \mathcal{R}$, which can be written as the contraction of factors
\begin{equation}
\rho=K_{\mathcal{L}} G K_{\mathcal{R}},
\end{equation}
where \(K_{\mathcal L}\) and \(K_{\mathcal R}\) correspond to the two grey-shaded contractions in Fig.~\ref{fig:tn-contraction}(d), and \(G\) is the contracted reduced part of the transition matrix.
We then perform LQ decomposition of the contracted $K_\mathcal{R}$, 
\begin{equation}
    K_{\mathcal R}=LQ, ~T=K_{\mathcal L}GL  
\end{equation}
with $T$ is a $D\times$ thinner matrix, with which we carry out the SVD truncation
\begin{equation}
    T=\widetilde U S W^\dagger, \rho=TQ=\widetilde U S W^\dagger Q=\widetilde U S\widetilde V^\dagger.
\end{equation}
Compared to the direct RTM truncation in \cite{carignanoTemporalEntropyComplexity2024a} that contracts all factors first and performs SVD on the full $\rho$, which costs $O(D^3 \chi^3)$, the above achieves a cost of $O(Dd\chi^3)$ by keeping the RTM in factorized form, hence giving a constant factor speedup in the algorithm.

As the RTM compression needs the environment as a compressed tMPS, in transverse contraction we first do an initial propagation using SRC from right to left in order to prepare all the right tMPS $\mathcal{R}_i$.
We then do a sweep from left to right to evolve all the left tMPS $\mathcal{L}_i$ with the initialized $\mathcal{R}_i$ as environments, and then sweep from right to left to update the $\mathcal{R}_i$ with $\mathcal{L}_i$ as environment. 

\noindent \emph{MPO evolution with SRC}. As a baseline method, we consider MPO evolution. This proceeds by starting from the middle of the tensor network where the observable is evaluated, and then applying the evolution gates as in Heisenberg evolution, in the form $\hat{O}(t+ dt) = U^\dag(dt) \hat{O}(t) U(dt)$ (see Fig.~\ref{fig:mpo-heisenberg}(a))~\cite{zwolakMixedStateDynamics2004,prosenOperatorSpaceEntanglement2007,hartmannDensityMatrixRenormalization2009}.
Note that the state-evolution quantum circuits end either with an evolution by $U_A$ (odd time-steps) or $U_B$ (even time-steps), thus exactly reproducing the state-evolution results in the Heisenberg picture requires two MPO trajectories (odd times starting with $U_A$, and even times starting with $U_B$). For simplicity, we only compute a single trajectory (starting with $U_B$), thus our odd-step results contain an additional small finite-$dt^2$ Trotter error, which we have included in the reported RMSE. 
Assuming $\hat{O}(t)$ is represented by an MPO of bond dimension $\chi$, then the exact $\hat{O}(t+dt)$ has bond dimension $D^2\chi=16 \chi$ and thus should be compressed. 
We adapt successive randomized compression for MPO-MPS products \cite{camanoSuccessiveRandomizedCompression2026a} to compress the MPO sandwich $U^\dag(dt) \hat{O}(t) U(dt)$. 
Rather than first fusing the three tensors at each site into an enlarged MPO, we attach each local random sketch directly to the pair of open physical legs of the sandwich and contract the resulting network successively, as shown in Fig.~\ref{fig:mpo-heisenberg}(b).
Hence the SRC performs the compression with cost $O(LDd \chi^3)$ compared to the $O(LD^3 d \chi^3)$ cost of the standard SVD compression algorithm. 

Note that for each time point, the forward and backward evolution gates outside the backward causal cone of the observable cancel exactly, allowing this part of the tensor network to be removed before contraction~\cite{banulsMatrixProductStates2009,hastingsConnectingEntanglementTime2015,friasPerezLightConeTensor2022}.

\noindent \emph{Evaluation of many observables}.
The two TN simulation methods have different cost-scaling when evaluating many observables, because they distribute the computation differently over time versus over the set of measured operators.
In MPO evolution, Heisenberg-evolving a single operator $\hat O$ from $t=0$ up to a final time $T$ passes through every intermediate Trotter step along the way, so one evolution run yields the entire time trace $\hat O(t)$, $t\in[0,T]$, of that observable at no extra cost.
However, a different observable, or observables on different sites, correspond to a different operator MPO with no work in common, so the cost of the MPO approach is cumulative in time {per observable}, and scales linearly with the number of distinct observables measured.
Each computation is independent, however, and thus may be executed on a separate resource.

For the transverse contraction, as Fig.~\ref{fig:tn-contraction}(a) shows, the contraction of one tensor network only gives the expectation value of one observable instead of the full description of the time evolved state.
However, a set of temporal MPS extracted from the identity TN $\langle \psi_0|U^\dagger(T) U(T)| \psi_0 \rangle$ (we call this temporal MPS propagation an identity sweep), with no observable operator during the evolution, can be reused to calculate observables.
In the length $L$ fermionic chain, given all the $2(L-1)$ left and right tMPS from the identity TN
\begin{equation}
    \label{eq:identity-environments}
    \mathcal{L}_i, \mathcal{R}_i, \qquad i=0,\ldots,L-2
\end{equation}
we can efficiently compute any local one- or two-site observable as a TN contraction of 3 or 4 temporal layers,
\begin{equation}
    \label{eq:local-observable-readout}
    \begin{aligned}
        \langle n_{i\sigma}\rangle &= \mathcal{L}_{i-1} \cdot \mathcal{E}^{(n_{i\sigma})}_i \cdot \mathcal{R}_i
        \\
        \langle n_{i\sigma}n_{i+1,\sigma'}\rangle &= \mathcal{L}_{i-1} \cdot \mathcal{E}^{(n_{i\sigma})}_i \cdot \mathcal{E}^{(n_{i+1,\sigma'})}_{i+1} \cdot \mathcal{R}_{i+1}
    \end{aligned}
\end{equation}
with $\mathcal{E}_i^{(n_{i\sigma})}$ the temporal MPO with the observable operator attached (Fig.~\ref{fig:tn-contraction}(b)). 

When calculating a general pair of operators (e.g. the nonlocal density operators $\langle n_{i\sigma} n_{j \sigma'}\rangle$ with $i<j$), we carry out an ``operator sweep'' by evolving new sets of left tMPSs $\mathcal{L}_{k}^{(n_{i\sigma})}$ from $k=i$ to $L-2$, with the first observable $n_{i \sigma}$ inserted in the tensor network (e.g. from the observable TN $\langle \psi_0|U^\dagger(T) \hat{n}_{i \sigma} U(T)| \psi_0 \rangle$). 
$\mathcal{L}_{k}^{(n_{i\sigma})}$ can be compressed by its right environment $\mathcal{R}_k$ obtained during the previous identity sweep, and we store a total of \(2(L-1)+L(L-1)\) tMPSs during the extra $2(L-1)$ sweeps of observable TN contraction
\begin{equation}
    \label{eq:operator-carrying-environments}
    \begin{aligned}
        \mathcal{L}_i, \mathcal{R}_i, & \quad i=0,\ldots,L-2,
        \\
        \mathcal{L}_k^{(n_{i\sigma})}, & \quad i=0,\ldots,L-2,\quad k=i,\ldots,L-2,\quad \sigma=\uparrow,\downarrow.
    \end{aligned}
\end{equation}
As Fig.~\ref{fig:tn-contraction}(b) shows, these tMPS environments can then be used to contract the expectation value of a nonlocal observable such as $\langle n_{i\sigma} n_{j \sigma'}\rangle$,
\begin{equation}
    \label{eq:nonlocal-observable-readout}
    \langle n_{i\sigma} n_{j \sigma'}\rangle= \mathcal{L}_{j-1}^{(n_{i\sigma})} \cdot \mathcal{E}^{(n_{j\sigma'})}_j \cdot \mathcal{R}_j
\end{equation}

\noindent \emph{GPU and \textsf{Quimb} implementation}.
Both contraction strategies were implemented using \textsf{Quimb} \cite{grayQuimbPythonPackage2018}, a tensor network and quantum many-body simulation library with support for tensor networks on arbitrary graphs and with native GPU acceleration. \textsf{Quimb} supports TN with symmetries, but because of the small bond dimensions used in the calculations (see below), we did not use any symmetry.
All production runs reported here were obtained using a single Nvidia H200 GPU with complex-double precision. In these runs, additional parallelism on a single GPU was exploited by concurrently running multiple independent processes using Nvidia's CUDA Multi-Process Service (CUDA MPS). 
\section{Results}
\label{sec:results}

\subsection{Advantage over quantum hardware}
\label{sec:advantageOverQuantum}

\begin{figure}
    \centering
    \includegraphics[width=1\linewidth]{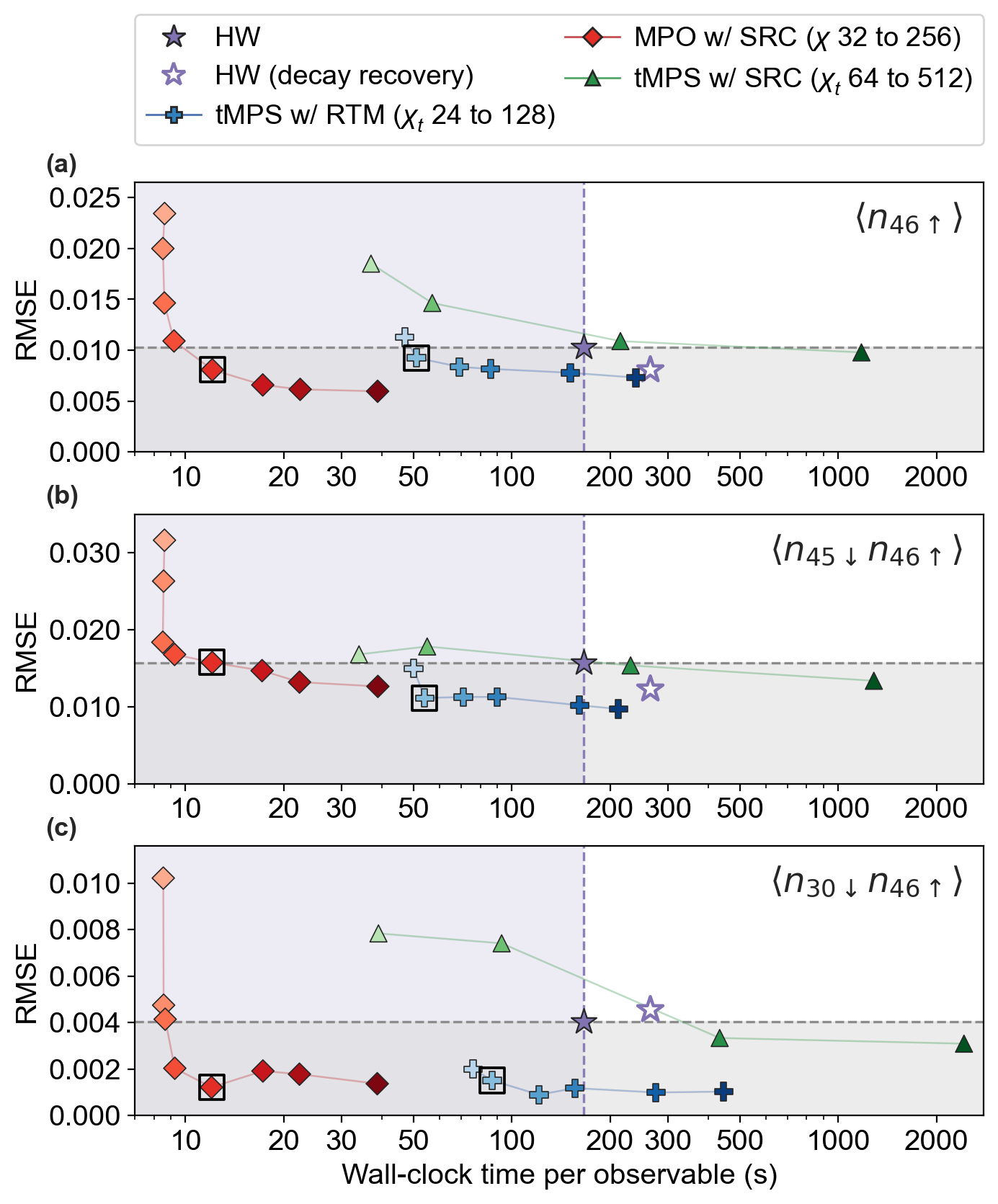}
    \caption{Cost-accuracy diagram for the three density observables of the $L=60$ Fermi-Hubbard Néel quench at $U/t_h=-2$. 
    Cost is defined by the total runtime to obtain the observable dynamics in $t\in [0,6]$ with Trotter step $dt=0.2$ on one QPU/H200 GPU.
    Accuracy is defined by the RMSE (Eq.~\ref{eq:rmse}) comparing with TDVP $\chi=4096$ data in $t\in [0,5.2]$.
    Bond dimension is encoded by color, light to dark as $\chi$ grows: blue crosses denote tMPS + RTM with $\chi_t = 24, 32, 48, 64, 100, 128$, red diamonds denote MPO + SRC with $\chi = 32, 48, 64, 96, 128, 160, 192, 256$, and green triangles denote tMPS + SRC with $\chi_t = 64, 128, 256, 512$. 
    Black squares highlight the MPO + SRC point at $\chi=128$ and the tMPS + RTM point at $\chi_t=32$ in each panel.
    The black square marks the datapoint we used in the runtime of Table \ref{tab:runtime}. 
    Filled and open purple stars denote HW and HW (decay recovery), respectively. The 166 s HW runtime includes the main circuits only, whereas the 265 s HW (decay recovery) runtime also includes the characterization circuits required for readout error mitigation and decay recovery~\cite{hartnettFastAccurateHighresolution2026a}.
    Purple shading marks runtimes below HW and grey shading marks RMSE below that panel's HW bar, so the doubly-shaded region is where a classical method beats the hardware on both axes at once. 
    (a) Runtime-RMSE of one-site observable $\langle n_{46 \uparrow}(t)\rangle$.
    (b) Runtime-RMSE of local two-site observable $\langle n_{45 \downarrow}(t) n_{46 \uparrow}(t)\rangle$. 
    (c) Runtime-RMSE of nonlocal two-site observable $\langle n_{30 \downarrow}(t) n_{46 \uparrow}(t)\rangle$.
    }
    \label{fig:cost-accuracy}
\end{figure}

\begin{figure}
    \centering
    \includegraphics[width=1\linewidth]{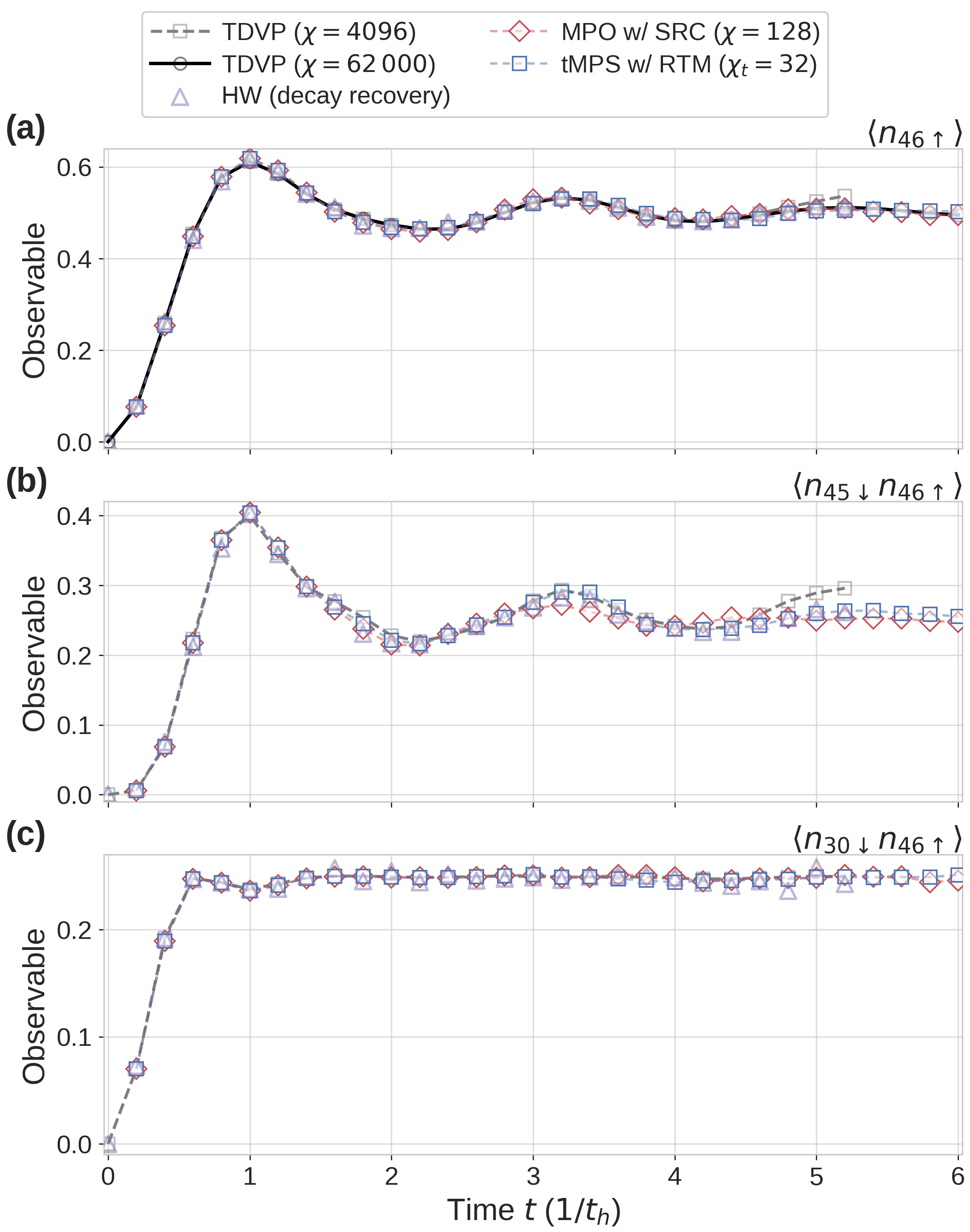}
    \caption{
     Dynamics of 3 observables from classical and quantum simulation. 
     Blue squares are tMPS w/ RTM at $\chi_t = 32$ and red diamonds are MPO w/ SRC at $\chi = 128$ from $t=0.2$ to $6.0$; lines connecting their sampling points are guides to the eye.
     Grey dashed squares are the TDVP $\chi=4096$ reference \cite{hartnettFastAccurateHighresolution2026a}, the black curve in panel (a) is the TDVP at $\chi_{\rm eff}\approx62\,000$ from \cite{rauschPushingClassicalFrontier2026}, and open purple triangles are HW (decay recovery) data, with readout error mitigation and decay recovery applied.
     (a) Dynamics of one-site observable $\langle n_{46 \uparrow}(t)\rangle$.
     (b) Dynamics of local two-site observable $\langle n_{45 \downarrow}(t) n_{46 \uparrow}(t)\rangle$. 
     (c) Dynamics of nonlocal two-site observable $\langle n_{30 \downarrow}(t) n_{46 \uparrow}(t)\rangle$.
    }
    \label{fig:dynamics}
\end{figure}

We first compare the cost and accuracy of computing a single observable's expectation value.
We consider the same three representative observables used in the benchmarks of Ref. \cite{hartnettFastAccurateHighresolution2026a}: the one-site density $\langle n_{46\uparrow}\rangle$, the nearest-neighbor
density correlation
$\langle n_{45\downarrow}n_{46\uparrow}\rangle$, and the spatially separated
correlation $\langle n_{30\downarrow}n_{46\uparrow}\rangle$.

Figure \ref{fig:cost-accuracy} reports the cumulative wall-clock time required to obtain each observable's dynamics over $0<t\leq6$ with $dt=0.2$ on one H200 GPU, with accuracy quantified by the RMSE against the TDVP $\chi=4096$ reference, with the error measured over the range $0<t\leq5.2$ (i.e. following the methodology in Ref.~\cite{hartnettFastAccurateHighresolution2026a} to define the hardware error)
\begin{equation}
\label{eq:rmse}
\begin{aligned}
\mathrm{RMSE}
&=
\sqrt{
\frac{1}{N_t}
\sum_{k=1}^{N_t}
\left[
\langle \hat O(t_k)\rangle
-
\langle \hat O(t_k)\rangle_{\mathrm{TDVP}}
\right]^2
},
\\
t_k&=0.2k,\quad N_t=26.
\end{aligned}
\end{equation}
The hardware results are shown by the filled and empty stars (without and with decay recovery and measurement mitigation), and classical results in the lower left quadrant indicate a lower error and faster speed than the quantum hardware, for a single observable.

In the MPO calculation, 
at $\chi=128$, all three observables reach better or similar accuracy to quantum hardware in only $\sim 12\mathrm{s}$, compared to the $\sim 1000$s for the MPO calculations in Ref.~\cite{hartnettFastAccurateHighresolution2026a}. The corresponding dynamics are shown in Fig.~\ref{fig:dynamics}.

For transverse contraction, we compare the environment-independent SRC compression (i.e. the warmup sweep in the transverse contraction) with the environment-informed RTM sweeps. 
The reported runtime in Fig.~\ref{fig:cost-accuracy} is the cumulative time for all the timepoints, where we parallelize over the independent timepoints (with at most 16 workers) on a single GPU.

Compared with SRC compression, the RTM sweeps provide an important improvement in bond-dimension efficiency.
Whereas tMPS+SRC reaches all three hardware RMSE thresholds only at $\chi_t=512$, tMPS+RTM reaches them already at $\chi_t=32$ with a shorter runtime; the resulting dynamics are shown in Fig.~\ref{fig:dynamics}. 
The observable timecost ranges from $\sim 50$s (local observables) to $\sim 90$s (non-local observables), as shown in Table \ref{tab:runtime}.

Note that in both the MPO and transverse contraction, increasing $\chi$ leads to a convergence of the RMSE error, but this converged value is not 0. 
One reason is because the $\chi=4096$ TDVP reference we are comparing to from Ref.~\cite{hartnettFastAccurateHighresolution2026a} also has a residual error from finite $\chi$ and an additional TDVP time-step error (which is different from Trotter error); using the data on single site observables from the more accurate TDVP simulations with $\chi\sim 62000$, we find the TDVP $\chi=4096$ RMSE to be $\sim$ 0.006 up to $t=5.2$.

\begin{table*}
\centering
\renewcommand{\arraystretch}{1.12}
\setlength{\tabcolsep}{3pt}
\resizebox{\textwidth}{!}{%
\begin{tabular}{l|ccc|cc|cc}
\hline
Method &
\shortstack{Single observable\\$\langle n_{46\uparrow}\rangle$} &
\shortstack{Single observable\\$\langle n_{45\downarrow}n_{46\uparrow}\rangle$} &
\shortstack{Single observable\\$\langle n_{30\downarrow}n_{46\uparrow}\rangle$} &
\shortstack{All one-site\\observables\\$\langle n_{i\mu}\rangle$} &
\shortstack{All nearest-neighbor\\two-site observables\\$\langle n_{i\mu}n_{i+1,\mu'}\rangle$} &
\shortstack{All pairs\\$\langle n_{i\mu}n_{j\mu'}\rangle$} &
\shortstack{All pairs (up to 120 GPUs)\\$\langle n_{i\mu}n_{j\mu'}\rangle$} \\
\hline
MPO+SRC ($\chi=128$) &
\textbf{$12.1\,\mathrm{s}$} &
\textbf{$12.1\,\mathrm{s}$} &
\textbf{$12.0\,\mathrm{s}$} &
$10.3\,\mathrm{min}$ & $19.9\,\mathrm{min}$ & $9.85\,\mathrm{h}$ & $\sim 4.93\,\mathrm{min}$ \\
tMPS+RTM ($\chi_t=32$) &
\textbf{$51\,\mathrm{s}$} &
\textbf{$54\,\mathrm{s}$} &
\textbf{$87\,\mathrm{s}$} &
\textbf{$122\,\mathrm{s}$} & $205\,\mathrm{s}$ & $54.3\,\mathrm{min}$ & \textbf{$\sim161\,\mathrm{s}$} \\
QPU  &
$166\,\mathrm{s}$ & $166\,\mathrm{s}$ & $166\,\mathrm{s}$ &
$166\,\mathrm{s}$ & $166\,\mathrm{s}$ & $166\,\mathrm{s}$ & $166\,\mathrm{s}$ \\
\hline
\end{tabular}%
}
\caption{Wall-clock time to obtain the indicated density observables for $t=0,0.2,\ldots,6.0$. Except for the last column, classical runtimes use one H200 GPU. The final column gives the estimated runtime with up to 120 H200 GPUs based on trivial parallelization; details of this estimation are given in Appendix~\ref{app:runtimeEst}.}
\label{tab:runtime}
\end{table*}

To compare the quantum and classical performance across different kinds of observables, Table~\ref{tab:runtime} reports the  wall-clock time required to obtain the observable dynamics for evolution from $t=0$ to 6.
The bare QPU execution time is 166s for every case because the same measured bitstrings provide all observables in the occupation basis. The time to obtain all observables differs between different classical TN contraction strategies, due to the different amount of reuse. 
For a single observable, MPO evolution is the current fastest classical approach ($\sim$12s for accuracy comparable to or better than the quantum hardware result).
Since each distinct observable needs a separate MPO simulation, the computational work is linear with the number of observables required. 
There are \(2L=120\) one-site densities and \(4(L-1)=236\) nearest-neighbor density correlators, so both local-observable workloads contain \({O}(L)\) independent MPO trajectories.
The complete set of two-point density correlations contains $7140$ observables,  requiring   \({O}(L^2)\) trajectories.
Computing all one-site densities, all two-site nearest-neighbour density correlators and all density pairs on a single GPU requires 10.3min, 19.9min, and 9.85h, as shown in Table \ref{tab:runtime}. 
(These times use 8 batched workers on the GPU to simultaneously compute multiple observables).

Transverse contraction instead reuses the \(2(L-1)\) identity TN environments \(\{\mathcal{L}_i,\mathcal{R}_i\}\) generated at each target time. As shown in Eq.~\eqref{eq:local-observable-readout}, all one-site and nearest-neighbor density observables can then be evaluated through small local contractions without additional tMPS sweeps.
Although constructing the environments of the full chain removes the lightcone savings available for a single observable, obtaining all 120 one-site densities and 236 nearest-neighbor correlators still requires only \(122\,\mathrm{s}\) and \(205\,\mathrm{s}\) of computation time on a single GPU, respectively.
(Here 12 workers were used on the GPU to parallelize over timepoints).

To compute all two-point density correlators $\langle n_{i \mu} n_{j \mu'} \rangle$, we additionally need the \(2(L-1)\) operator sweeps and to combine the operator-carrying left environments with the shared identity right environments according to Eq.~\eqref{eq:nonlocal-observable-readout}.
This construction 
produces all 7140 non-local correlators in \(54.3\,\mathrm{min}\) (with 8 batched workers) on a single GPU. 
This single GPU runtime is below the \(1\,\mathrm{h}\,40\,\mathrm{min}\) GPU-TDVP result reported in Ref.~\cite{rauschPushingClassicalFrontier2026} with comparable accuracy, although TDVP produces a full-wavefunction approximation rather than a specified set of density correlators.

Finally, we have estimated the cost of transverse contraction that uses a trivial parallelization over independent operator sweeps (see Appendix~\ref{app:runtimeEst}). Using up to 120 GPUs (i.e. 15 cloud computing instances, which is easily obtained), the estimated classical runtime then becomes less than the QPU runtime (161s, see Table \ref{tab:runtime}). Using commercial pricing (at the time of writing), the calculation corresponds to approximately USD 30 of compute. 

\subsection{Longer time dynamics}
\label{sec:longerTime}

\begin{figure}
    \centering
    \includegraphics[width=1\linewidth]{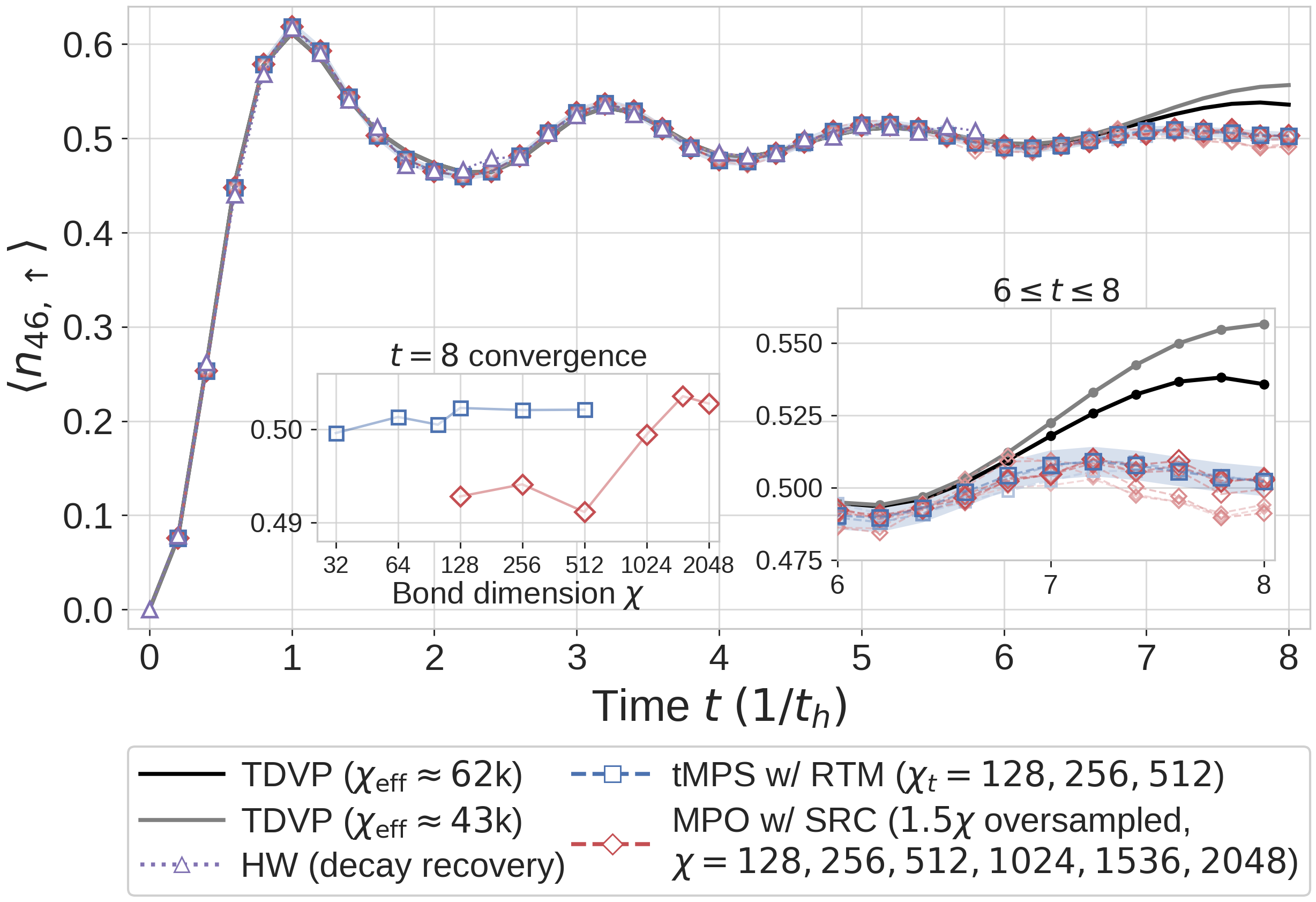}
    \caption{
    Longer-time dynamics of observable $\langle n_{46 \uparrow}\rangle$ over $0 \leq t \leq 8$. The right inset shows the late-time window $6\leq t\leq8$, while the left inset shows the bond-dimension convergence at $t=8$, with $\chi$ on a logarithmic axis.
    Reference TDVP at $\chi_{\rm eff}\approx62$k (black) and $\approx43$k (grey) with truncation cutoff $\epsilon_{\mathrm{tol}}=1\times 10^{-6}$ are solid lines, and the HW (decay recovery) data are purple triangles.
    The classical TN results are tMPS with RTM at $\chi_t = 128, 256, 512$ (blue squares) and MPO with SRC at $\chi = 128, 256, 512, 1024, 1536, 2048$ using an oversampled sketch dimension of $1.5\chi$ followed by retruncation to $\chi$ (red diamonds). 
    The $t=8$ convergence inset additionally includes tMPS with RTM at $\chi_t=32,64,100$.
    Light-blue shading marks a $\pm1\%$ band around tMPS+RTM at $\chi_t=512$.
    }
    \label{fig:longtime}
\end{figure}

Having established the accuracy and efficiency of our methods for \(t\leq6\), we now extend the simulation to longer times. 
Figure~\ref{fig:longtime} shows the local density \(\langle n_{46\uparrow}(t)\rangle\) computed up to \(t=8\), with larger bond dimensions for both tensor network methods.
After the initial transient, the density exhibits damped oscillations around half filling. 
For both MPO with SRC and tMPS with RTM compression, the results at the largest bond dimension overlap to within 0.73\%. From the left inset, we see that the tMPS result is likely converged to beyond this accuracy.

In contrast, the two largest TDVP calculations agree at earlier times but begin to separate beyond \(t\approx7\), despite reaching effective bond dimensions of approximately \(43{,}000\) and \(62{,}000\) with a small cutoff of $\epsilon_{\mathrm{tol}}=1\times 10^{-6}$~\cite{rauschPushingClassicalFrontier2026} (note, this analysis is based on the results presented in the SI of Ref.~\cite{rauschPushingClassicalFrontier2026}, rather than the results with larger cutoff of $\epsilon_{\mathrm{tol}}=1\times 10^{-4}$ shown in Fig.~1 of Ref.~\cite{rauschPushingClassicalFrontier2026}). 
The methods we have used therefore can access the dynamics of this local observable beyond a time-scale where representing the complete many-body wavefunction is classically difficult. 

\FloatBarrier
\section{Conclusion and Discussion}
\label{sec:discussion}

We have shown that the observable dynamics of the 60-site/120 qubit Fermi--Hubbard quench simulated on quantum hardware in Ref.~\cite{hartnettFastAccurateHighresolution2026a} can be computed classically using MPO evolution and transverse contraction, and that with transverse contraction, the task of computing all reported 7260 observables can be completed faster than the quantum runtime, with moderate resources and with minimal cost (approximately 30 USD of accelerator time). 

It is likely that, with further modest effort, the reported classical runtimes can be reduced further (see Appendix \ref{app:runtimeEst}). The accounting of the quantum processor cost is already quite favourable, as no classical pre/postprocessing cost is included. With the given circuit choice, the only way to improve the quantum runtime would be to increase the gate execution speed. Alternatively, moving to a fault-tolerant quantum architecture to perform the identical experiment, while leading to a significant slowdown in running the circuit due to error correction, would likely yield higher accuracy results (assuming the errors here are not entirely shot limited). Reproducing such higher accuracy would require additional classical cost.

Classical computers are not expected to be able to simulate arbitrary quantum dynamics efficiently, and assuming this, asymptotic theoretical quantum advantage in quantum dynamics is guaranteed. But practical quantum advantage is about speedup for specific problem instances and tasks. The methods used here return observable trajectories rather than the full quantum state, a more natural formulation for classical computation. However, when this is the assigned task in modeling the short-time dynamics of the 1D Fermi-Hubbard model, 
our work shows that classical methods are a competitive and practical choice.

\section{Acknowledgments}

This work was primarily supported by the US Department
of Energy, Office of Science, Accelerated Research in
Quantum Computing Centers, Quantum Utility through
Advanced Computational Quantum Algorithms, through
Award No. DE-SC0025572. GKC acknowledges support from the Institute for Quantum Information and Matter.
\FloatBarrier
\bibliography{refs}
\clearpage
\onecolumngrid
\renewcommand{\topfraction}{0.95}
\renewcommand{\bottomfraction}{0.95}
\renewcommand{\textfraction}{0.05}
\renewcommand{\floatpagefraction}{0.85}
\appendix
\counterwithin{figure}{section}
\renewcommand{\thefigure}{\thesection\arabic{figure}}

\FloatBarrier
\section{Runtime estimate for all 7260 observables}
\label{app:runtimeEst}

Here we explain how the multi-GPU runtimes in the final column of Table~\ref{tab:runtime} are estimated.

For MPO+SRC, the operator evolution for every observable is independent. 
Assuming an even distribution of the trajectories (each one evolves from $t=0$ to $6$), the measured single GPU runtime can therefore be divided among 120 GPUs. 
This gives an estimated runtime of $9.85\,\mathrm{h}/120=4.93\,\mathrm{min}$ in the final column of Table~\ref{tab:runtime}.

For tMPS+RTM, the final column of Table~\ref{tab:runtime} is a conservative upper-bound estimate. 
To calculate correlation function $\langle n_{i \mu} n_{j \mu'} \rangle$, we batch the calculation into $2L=120$ jobs indexed by $(i,\mu)$ and assign one job to each of 120 GPUs.
Each GPU evaluates the 30 target times $t=0.2,0.4,\ldots,6.0$ using 16 persistent workers using CUDA MPS. 
Each job has two phases. Phase 1 is the identity sweep. 
It consists of an SRC initialization followed by two RTM sweeps and constructs the identity tMPSs $\{\mathcal{L}_k,\mathcal{R}_k\}$ in Eq.~\eqref{eq:identity-environments} for every target time. 
Phase 2 is the operator sweep.
In the job indexed by $(i,\mu)$, the operator sweep inserts $n_{i\mu}$ into the temporal column at site $i$, as shown in Fig.~\ref{fig:tn-contraction}(b), and propagates the operator-carrying left tMPS $\mathcal{L}^{(n_{i\mu})}_k$ towards the right using the cached $\mathcal{R}_k$ as the RTM compression environment according to Eq.~\eqref{eq:operator-carrying-environments}. 
At every site $j>i$, both spin choices of the second operator are evaluated using Eq.~\eqref{eq:nonlocal-observable-readout}, yielding $2(L-1-i)$ intersite correlations. 
The one-site occupation $\langle n_{i\mu}\rangle$ and double occupation $\langle n_{i\uparrow} n_{i \downarrow}\rangle$ are also read out at the starting site. 
The 118 nonempty operator sweeps then cover all 7080 intersite correlations.
The remaining local readouts are obtained from the identity tMPSs using Eq.~\eqref{eq:local-observable-readout}. 
Together, the 120 jobs yield all 120 one-point and 7140 two-point density observables.

\begin{table}
\centering
\renewcommand{\arraystretch}{1.12}
\setlength{\tabcolsep}{3pt}
\resizebox{\textwidth}{!}{%
\begin{tabular}{c|c|c|c|c|c|c}
\hline
Carried operator &
\shortstack{Number of compressions\\in operator sweep} &
\shortstack{Number of\\observables} &
\shortstack{Identity sweep\\runtime} &
\shortstack{Operator sweep and\\readout runtime} &
\shortstack{Other overhead\\runtime} &
\shortstack{Total\\runtime} \\
\hline
$n_{0\uparrow}$ & 58 & 120 & $97.3\,\mathrm{s}$ & $56.9\,\mathrm{s}$ & $6.8\,\mathrm{s}$ & $161\,\mathrm{s}$ \\
$n_{30\downarrow}$ & 29 & 60 & $98.1\,\mathrm{s}$ & $28.6\,\mathrm{s}$ & $5.3\,\mathrm{s}$ & $132\,\mathrm{s}$ \\
$n_{45\downarrow}$ & 14 & 30 & $97.8\,\mathrm{s}$ & $14.2\,\mathrm{s}$ & $6.0\,\mathrm{s}$ & $118\,\mathrm{s}$ \\
$n_{46\uparrow}$ & 13 & 28 & $97.4\,\mathrm{s}$ & $13.0\,\mathrm{s}$ & $6.6\,\mathrm{s}$ & $117\,\mathrm{s}$ \\
\hline
\end{tabular}%
}
\caption{Measured runtime decomposition for four representative jobs $(0, \uparrow), (30, \downarrow), (45, \downarrow), (46, \uparrow)$ at $\chi_t=32$. The runtime of the identity sweep and operator sweep is obtained using 16 CUDA-MPS workers on one H200 GPU. 
The number of observables includes the insertion-site occupation and all $j>i$ two-point correlations returned by that job.}
\label{tab:allpairs-runtime-breakdown}
\end{table}

Table~\ref{tab:allpairs-runtime-breakdown} shows that the runtime of the identity sweep is roughly independent of the site $i$ of operator insertion. 
The operator sweep uses less time as the first insertion moves to the right and fewer operator-carrying tMPS propagation steps are required. 
The first-site job with $n_{0\uparrow}$ inserted is therefore the longest job. Since the parallelism over the 120 GPUs is over \emph{completely independent} operator sweeps (although each takes a different amount of time), we can take the first-site job runtime to be representative of (and in practice, upper bound, see entry $n_{0\uparrow}$ in Table~\ref{tab:allpairs-runtime-breakdown}) the time for all the operator sweeps. This gives the conservative multi-GPU runtime estimate of $161\,\mathrm{s}$. 

The corresponding cost is based on the cloud compute pricing at time of writing \cite{amazonEC2CapacityBlocksPricing2026}, $5.721$ USD per H200-hour. This gives the estimated GPU cost
\begin{equation}
    120\times\frac{161\,\mathrm{s}}{3600\,\mathrm{s}}
    \times 5.721\ \mathrm{USD}
    \simeq 30.7\ \mathrm{USD}.
\end{equation}

This estimation retains some conservative features that could be improved to save time and resources.
First, the operator sweep begins only after the identity sweep has finished for all 30 target times, although each target time is independent and can proceed immediately once its own identity environments are available.
Second, the same identity sweep is repeated on every GPU, even though $\{\mathcal{L}_k,\mathcal{R}_k\}$ is common to all choices of $(i,\mu)$ at a fixed timepoint.
Computing these environments once and distributing them to the operator-sweep workers would reduce the repeated work. 
Also, as  Table \ref{tab:allpairs-runtime-breakdown} shows, multiple jobs with shorter operator sweeps could also share one GPU, since shorter sweeps complete in less than  161s.
These alternatives would reduce the required GPU count or total GPU time, although, unlike for the simple strategy above, an accurate time estimation would require additional implementation to verify.

\end{document}